# Radical-mediated Electrical Enzyme Assay For At-home Clinical Test

Hyun-June Jang[1,2]‡, Hyou-Arm Joung[3]‡, Xiaoao Shi[1,2], Rui Ding[1,2], Justine Wagner[4,5,6], Erting Tang[1], Wen Zhuang[1,2], Byunghoon Ryu[7], Guanmin Chen[8], Kiang-Teck Jerry Yeo[8,9], Jun Huang[1], Junhong Chen[1,2]*

[1]Pritzker School of Molecular Engineering, University of Chicago, Chicago, IL 60637, USA

[2]Chemical Sciences and Engineering Division, Physical Sciences and Engineering Directorate, Argonne National Laboratory, Lemont, IL 60439, USA

[3]Kompass Diagnostics Inc., Lombard, IL 60148, USA

[4]School of Chemistry and Biochemistry, Georgia Tech Polymer Network, Georgia Institute of Technology, Atlanta, Georgia 30332, United States

[5]School of Materials Science and Engineering, Georgia Tech Polymer Network, Georgia Institute of Technology, Atlanta, Georgia 30332, United States

[6]Center for Organic Photonics and Electronics, Georgia Tech Polymer Network, Georgia Institute of Technology, Atlanta, Georgia 30332, United States

[7]Department of Mechanical Engineering, Inha University, Incheon, 22212, Republic of Korea

[8]Department of Pathology, The University of Chicago, Chicago, IL 60637, USA

[9]Pritzker School of Medicine, The University of Chicago, Chicago, IL 60637, USA

‡These authors contributed equally to this work

*Correspondence: junhongchen@uchicago.edu

**Keywords:** field-effect transistor biosensors, paper-based fluidics, immunoassay, estradiol, proton sensing, radical-mediated electrical enzyme assays

**Abstract**

To meet the growing demand for accurate, rapid, and cost-effective at-home clinical testing, we developed a radical-mediated enzyme assay (REEA) integrated with a paper fluidic system and electrically read by a handheld field-effect transistor (FET) device. The REEA utilizes horseradish peroxidase (HRP) to catalyze the conversion of aromatic substrates into radical forms, producing protons detected by an ion-sensitive FET for biomarker quantification. Through screening 14 phenolic compounds, halogenated phenols emerged as optimal substrates for the REEA. Encased in an affordable cartridge ($0.55 per test), the system achieved a detection limit of 146 fg/mL for estradiol (E2), with a coefficient of variation (CV) below 9.2% in E2-spiked samples and an r² of 0.963 across a measuring range of 19 to 4,551 pg/mL in clinical plasma samples, providing results in under 10 minutes. This adaptable system not only promises to offer a fast and reliable platform, but also holds significant potential for expansion to a wide array of biomarkers, paving the way for broader clinical and home-based applications.

Healthcare systems worldwide are struggling to meet the demands of aging populations and increasingly complex health needs, compounded by workforce shortages, capacity constraints, and rising costs.[1,2] In this context, home-based care has gained prominence, offering high patient satisfaction, cost-effective management, and reduced readmission rates, all while maintaining healthcare quality.[3] The shift toward home-based healthcare has underscored the urgent need for accurate, low-cost, rapid, and portable point-of-care (POC) clinical testing to manage a variety of conditions.[4-6]

A particularly relevant application is monitoring fertility hormones.[7-9] High estradiol (E2) levels can contribute to conditions such as fibroids, endometriosis, and increased cancer risk.[10] Conversely, low E2 levels may indicate ovarian insufficiency or menopause, impacting fertility and reproductive health.[11] However, hormone assays for estrogen usually need lab-based chemiluminescence immunoassays due to its low and variable concentration during pregnancy (20 pg/mL to 20,000 pg/mL) and susceptibility to biomolecular interference.[12] Crucially, extremely low concentrations of E2 under a few pg/mL, relevant for non-reproductive issues, challenge current analytical methods due to difficulties in achieving accurate measurements.[13]

On the one hand, incorporating CLIA techniques into portable devices is fraught with challenges due to the unstable nature of reagents, necessitating sophisticated microfluidic systems and optical detectors that escalate both complexity and cost.[14] On the other hand, field-effect transistor (FET) biosensors, while providing rapid and sensitive fg/mL level detection[15,16] demand complex fluidic controls to address sample matrix interference[17,18] and Debye length constraints[16,17]. Maintaining the necessary wet interface for FETs also introduces risks of fluid leakage and damage, complicating commercial viability. Furthermore, consistent control of batch-to-batch and device-to-device variability is essential for broad market adoption.[17]

In clinical settings, horseradish peroxidase (HRP) catalyzes substrate transformations, generating optical signals directly linked to analyte concentrations.[19] However, by repurposing HRP-substrate interactions for electrical detection, protons generated during the reaction can be quantified by a FET, with proton count correlating to analyte concentration. This approach, termed radical-mediated electrical enzyme assay (REEA), overcomes sample matrix interference, allowing for the direct use of human plasma samples and enabling a simplified fluidic design while maintaining high sensitivity and a miniaturized device.

Here, we present a diagnostic platform that integrates REEA with a paper-based fluidic system[20] and a handheld FET reader. The FET quantifies protons generated within the paper fluidics cartridge by the REEA system, correlating the proton count with the concentration of target analytes. As a proof-of-concept, estradiol (E2), a key fertility hormone, was measured with a limit of detection (LOD) of 146 fg/mL under E2-spiked buffer condition and a coefficient of variation below 9.2%, using a cartridge costing $0.55 per test. The system demonstrated a strong correlation ($r^2$ = 0.963) across a measuring range of 19 to 4,551 pg/mL for 23 plasma samples, with a standard error of estimate (Sy/x) of 0.15 in a logarithmic regression analysis. These results are comparable to those obtained with an FDA-cleared clinical immunoassay on the Cobas e801 analyzer, indicating high accuracy. Screening 14 phenolic compounds interacting with HRP identified halogenated phenolic substrates—such as 4-fluorophenol (FP), 4-iodophenol (IP), 4-chlorophenol (CP), and 4-bromophenol (BP)—as the most effective for REEA. Additionally, the REEA cartridge was equipped with an enzymatic choline oxidase (ChOx)-based system for automated hydrogen peroxide ($H_2O_2$) generation. This integration of FET technology with paper fluidics and the REEA system offers a low-cost, accurate, and accessible solution for at-home clinical test, with broad potential for detecting a wide range of biomarkers across various clinical applications.

**RESULTS**

**Detection Platform**. Figure 1a depicts the handheld FET reader device (dimensions 2.5 cm by 4 cm) and the paper-based immunoassay cartridge for automated $H_2O_2$ generation system. The cost of cartridge is $0.55 per test (Table S1). To operate, the clinical sample is initially mixed with optimized running buffer and incubated for 15 minutes (Figure 1b). Following incubation, the cartridge, which includes the sensing electrodes such as indium tin oxide (ITO) and reference electrodes, is electrically connected to designated terminals on the handheld device (Figure S1). A total 300 µL of mixture of plasma and reagents is then injected into the cartridge, and the device begins real-time measurement of the signal over a 10-minute period (*see* Experimental Section in Supporting Information).

**REEA Mechanism**. HRP, an enzyme traditionally used in analytical techniques to convert substrates into optical signals (Figure 1c), is repurposed in the REEA system to convert aromatic substrates into radical forms, generating protons[21,22] (Figure 1d). HRP, which contains an Fe(IV) oxyferryl center, undergoes oxidation by $H_2O_2$, leading to the formation of compound I.[23] This compound I then interacts with a phenol compound (PhOH) at its active site, generating and releasing a phenolic free radical and a proton into the medium. The subsequent reaction produces compound II, which further reacts with another phenol molecule, releasing an additional proton. These protons, which correlate with the concentration of the target analyte, serve as the signal in the REEA system and are quantified by an ion-sensitive FET (Figure 1e). The intrinsic pH sensitivity of the ITO electrode, measured using a remote-gate FET (RGFET)[18,24] setup (Figure S2a), exhibited a proton-specific Nernstian response of 52.4 mV/pH with an $r^2$ of 0.998 (Figure S2b), showing no interference from varying surface areas of the testing solution on the ITO due to high input impedance of FET[25] (Figure S2c).

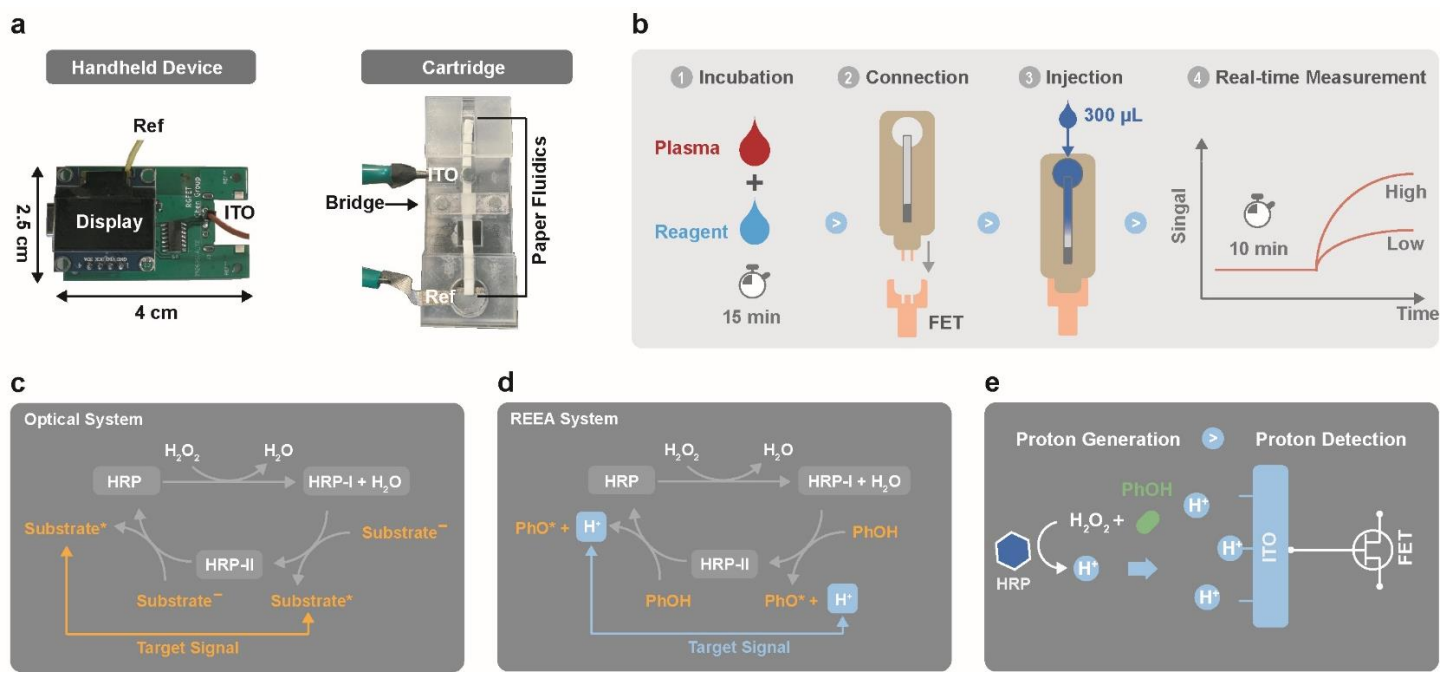

**Figure 1.** (a) Images of the handheld device and cartridge, alongside (b) schematic diagrams illustrating the operational workflow. The signal generation mechanisms by HRP and substrate reaction are shown for (c) conventional optical systems, (d) the REEA system, and (e) proton detection as a result of the REEA reaction via FET.

A set of phenolic compounds with a wide range of pKa values (Figure 2a), such as 2,4,6-trimethoxyphenol (TMYP), 2,4,6-trimethylphenol (TMLP), 4-methoxyphenol (MP), FP, IP, CP, BP, 2,4,6-trichlorophenol (TCP), p-anisidine (AD), p-toluidine (T), 4-fluoroaniline (FA), o-phenylenediamine (OPD), 2,4,6-trimethylaniline (TMYA), and 4-chloroaniline (CA), were screened to determine the optimal substrate for REEA in the solution phase through the RGFET setup (Figure S3a). Different degrees of shifts in threshold voltage ($V_{th}$) were observed in Figure 2b by the reaction between HRP, $H_2O_2$, and each substrate (Figure S3b, S3c). Without $H_2O_2$ and HRP (Figure S3d), and with either $H_2O_2$ (Figure S3e) or HRP (Figure S3f) alone, changes in $V_{th}$ were minimal for all substrates.

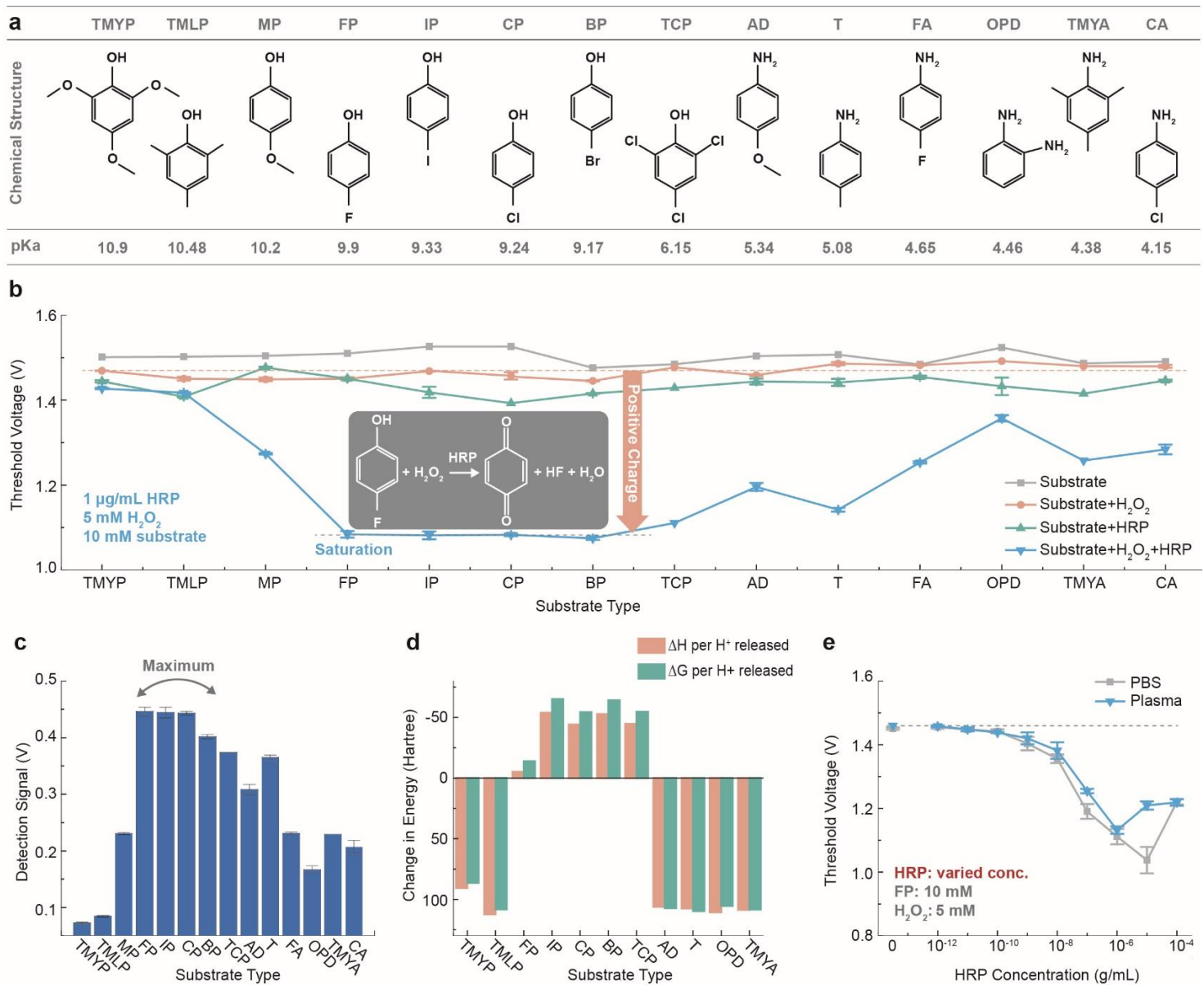

**Figure 2.** (a) The chemical structures of the phenolic compounds tested for the REEA system. (b) The distribution of $V_{th}$ versus substrate reacted with HRP and $H_2O_2$, including controls with substrate alone, substrate with $H_2O_2$ but no HRP, and substrate with HRP but no $H_2O_2$. (c) Detection signals as a function of substrate type. (d)

Changes in ΔH and ΔG simulated by DFT for each substrate type. (e) $V_{th}$ versus HRP concentration at 10 mM FP and 5 mM $H_2O_2$ diluted in either PBS or plasma.

The detection signal, defined as the difference in $V_{th}$ between the HRP/$H_2O_2$/substrate mixture and the substrate-only condition, reveals that halogenated phenolic compounds—specifically FP, IP, CP, and BP—generated the most substantial signals (Figure 2c). The oxidation of these compounds results in the release of halogenated ions[26,27], leading to the formation of hydrofluoric acid (HF), hydroiodic acid (HI), hydrochloric acid (HCl), and hydrobromic acid (HBr), respectively (inset of Figure 2b). These strong acids contribute to a significant reduction in $V_{th}$. Density functional theory (DFT) simulations[28] conducted to calculate changes in enthalpy (ΔH) and Gibbs free energy (ΔG) supports that proton generation is highly favorable for these halogenated ions (Figure 2d, Table S2). Considering the solubility of the substrates in aqueous solutions, FP was selected for subsequent experiments.

The REEA system effectively addresses Debye length issues, as demonstrated by significant $V_{th}$ shift in both 1X phosphate-buffered saline (PBS) (Figure S4a) and plasma (Figure S4b) with HRP concentrations ranging from 10 pg/mL to 1 µg/mL (Figure 2e). However, at HRP concentrations exceeding 10 µg/mL, a reversal in $V_{th}$ was observed for both PBS and plasma, suggesting a potential saturation point in the REEA signal. This reversal is likely due to HRP inactivation (Figure S4a, S4b), where the disruption of the heme macrocycle by phenoxyl radical attacks impairs its catalytic function.[29]

**Validation of Handheld Reader System.** The real-time change in the drain current of the FET, measured at a constant gate voltage of 2 V in the handheld device (Figure S5a), results from alterations in surface potential on the ITO and is converted to an output voltage ($V_{out}$). The real-time pH sensitivity (Figure S5b), measured by the $V_{out}$ change of the handheld FET reader in response to pH of testing solution, exhibited an amplified response of 212 mV/pH, further enhanced by an operational amplifier built in the handheld device, demonstrating 99.4% linearity (Figure S5c) and negligible drift (Figure S5d). Potential interferences caused by various paper materials, including nitrocellulose (NC) membrane, cotton linter pad, cellulous filtration paper, two glass fiber pads, and polysulfone membrane, inserted between the ITO electrode and the solution (Figure S6a), were found to be insignificant. That is, the $V_{out}$ signal remained specific to the pH of the buffer solution, regardless of the paper type connected to the ITO electrode (Figure S6b). The handheld FET reader's detection performance, as evaluated by the HRP-FP reaction in solution phase (Figure S7a), closely matched that of a precise semiconductor analyzer (Figure S7b).

A copper (Cu) reference electrode was used as an alternative to the standard Ag/AgCl reference electrode in the miniaturized cartridge system (Figure S8). The Cu reference electrode demonstrated pH sensitivity comparable to that of the Ag/AgCl electrode (Figure S8b). Notably, the signal trend in HRP-FP reaction measured by the Cu electrode (Figure S8c) closely matched the signal from the Ag/AgCl electrode (Figure S8d), as shown in Figure S8e.

**Validation of REEA within paper fluidics.** To demonstrate the high sensitivity of REEA system within the paper fluidic system, the LOD for both optical and REEA systems was compared directly with a conventional lateral flow assay format. The NC membrane, dried with varying concentrations of anti-E2 antibody, was tested by adding anti-mouse IgG-HRP. After washing step, the strips were imaged using a tetramethylbenzidine (TMB) solution, producing distinct bands corresponding to different anti-E2 antibody concentrations (inset of Figure 3a). The LOD measured by colorimetry was estimated to be 80 ng/mL (Figure 3a). However, when the same paper fluidics setup was characterized using the HRP-FP reaction, the LOD was determined to be at least 1,000 times lower than that achieved through colorimetry (Figure 3a, S9).

To enhance the HRP signal for target analyte binding, gold nanoparticles (AuNPs) were conjugated with estradiol-HRP (E2-HRP), designed for the E2 detection system. The AuNP-HRP-E2 conjugates were dried onto the NC membrane and tested across a concentration range of 0 to 10 optical density (OD at 525 nm) (Figure S10a). The HRP-FP reaction produced an intense signal even at a low AuNP-HRP-E2 concentration of 0.025 OD (Figure S10b), demonstrating that AuNPs conjugated with HRP function effectively within the REEA system, as also shown in Figure 3a.

$H_2O_2$, a key component in REEA, is unstable under ambient conditions, making automated $H_2O_2$ generation upon sample injection critical for ensuring long-term stability and commercial scalability of the system. This was achieved through the interaction between ChOx and choline chloride (CC) in the signal zone (Figure S1) of the cartridge. The ChOx-CC enzymatic reaction produces $H_2O_2$, which triggers the secondary enzymatic reaction between HRP and the REEA substrate (Figure 3b). To validate this automated $H_2O_2$ generation system,

ChOx was dried onto the NC membrane (inset of Figure 3c) and tested with an injected solution containing AuNP-HRP-E2 conjugate (0.05 OD), 25 mM CC, and 10 mM of each substrate identified through screening (Figure 2). The resulting REEA signals closely matched those observed in Figure 2b, while negligible signals were detected in negative control 1 (N1: no substrate, with HRP and CC), negative control 2 (N2: no substrate and HRP, with CC), and the blank control (buffer only). These results confirm the system's specificity and functionality, demonstrating its effectiveness in automating $H_2O_2$ generation for the REEA process.

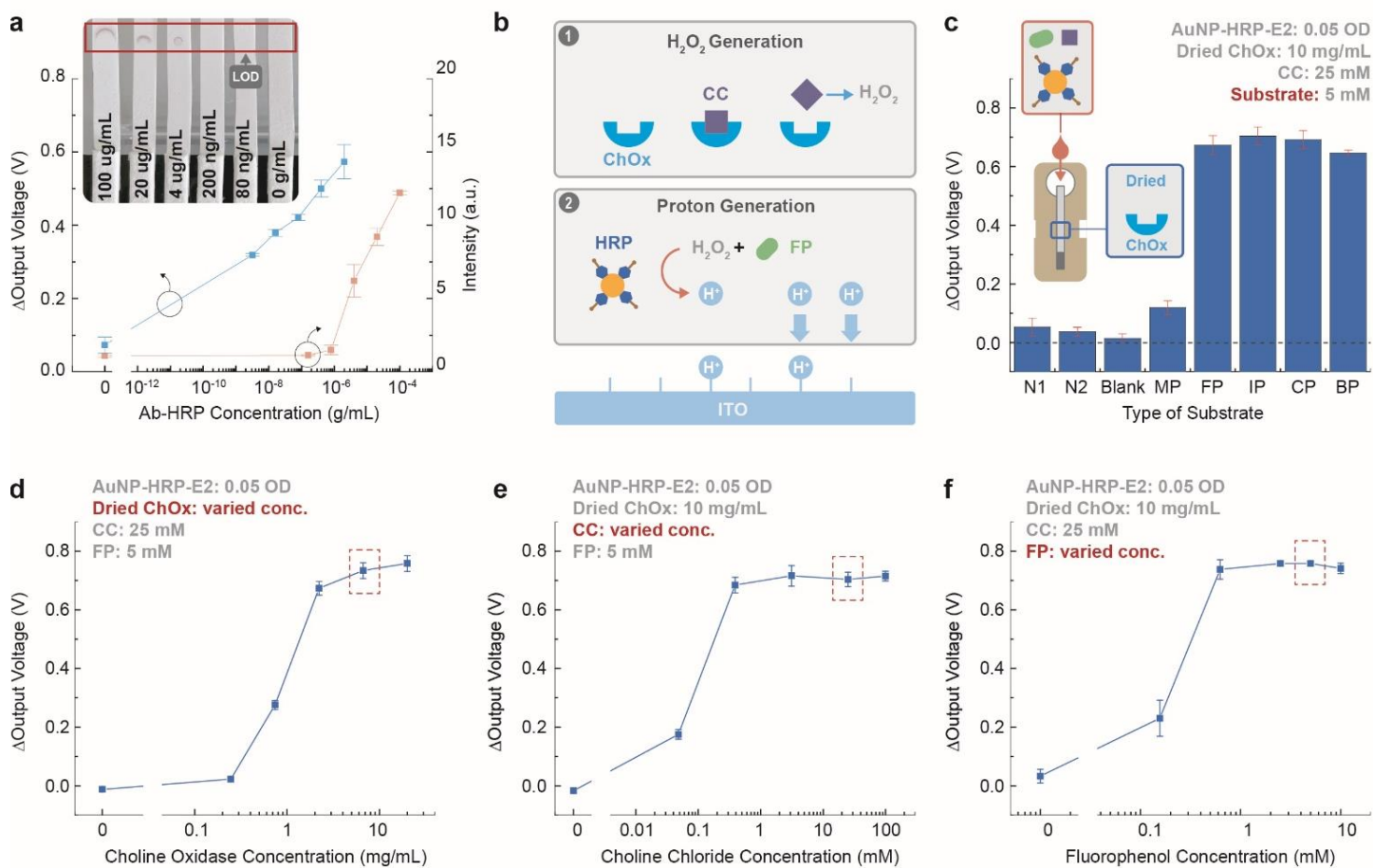

**Figure 3.** (a) Signal comparison from paper fluidics strips spotted with anti-E2 antibodies (0 to 100 μg/mL), analyzed using the REEA system and the handheld device, alongside colorimetry. The inset shows images of paper fluidics strips characterized using TMB. (b) Schematic of the automated $H_2O_2$ generation system, illustrating the interaction between ChOx and CC and the subsequent HRP-FP reaction. (c) $\Delta V_{out}$ of paper fluidics dried with ChOx and analyzed with AuNP-HRP-E2, CC, and each REEA substrate. The inset shows components of the cartridge and the testing solution. (d) $\Delta V_{out}$ of paper fluidics dried with varying concentrations of ChOx (for testing solution: AuNP-HRP-E2 at 0.05 OD, CC at 25 mM, and FP at 5 mM). (e) $\Delta V_{out}$ of paper fluidics dried with 10 mg/mL ChOx in response to different CC concentrations (for testing solution: AuNP-HRP-E2 at 0.05 OD, and FP at 5 mM). (f) $\Delta V_{out}$ of paper fluidics dried with 10 mg/mL ChOx in response to varying FP concentrations (for testing solution: AuNP-HRP-E2 at 0.05 OD, and CC at 25 mM).

Using the same testing setup as shown in Figure 3c, the concentration of ChOx dried on the NC membrane was optimized for the REEA system to ensure sufficient $H_2O_2$ production without compromising the REEA signal (Figure 3d). A ChOx concentration above 2 mg/mL resulted in a saturated response (Figure S11a), with 10 mg/mL selected as the optimal concentration for subsequent experiments. Similarly, the concentration of CC in the injected solution was varied while maintaining constant levels of dried ChOx (10 mg/mL), AuNP-HRP-E2 (0.05 OD), and FP (5 mM) (Figure 3e, S11b). Saturation was observed at CC concentrations above 1 mM, with 25 mM chosen as the optimal concentration. Finally, FP concentration was optimized under the conditions established in Figure 3d and 3e. FP concentrations above 0.5 mM led to signal saturation, and 5 mM was selected as the optimal concentration for further experiments (Figure 3f, S11c).

In the next step, surfactants for the testing solution were screened in Figure 4a by evaluating 24 different surfactants listed in Table 3. The investigation focused on E2 capture efficiency, signal resolution, and proton generation efficiency, with each surfactant mixed into a 2% BSA in 1x PBS solution. IGEPAL was selected for its ability to increase E2 capture efficiency by 21.8% and proton generation efficiency of REEA system by 31.06%, while maintaining good baseline control (CV: 3.04%). However, IGEPAL slightly reduced the signal resolution by 12.93%, where signal resolution is defined as the difference between the signal for a sample with no free E2 and the signal for the same solution spiked with 1 ng/mL of E2. (Table S3). To mitigate this, stabilizer components were screened in Figure 4b, leading to the selection of 2.5% PVP, which enhanced the signal resolution by 66.49%. The final optimized reagent comprised 2% BSA, 0.2% IGEPAL, and 2.5% PVP.

**REEA cartridge for E2 Detection.** Cartridge operation begins with the incubation of the testing sample, as shown in step 1 of Figure 1b, where a competitive binding reaction occurs: E2 antibodies bind either to AuNP-HRP-E2 or free E2 (inset of Figure 4c). Due to its smaller size, free E2 has a higher binding affinity for the antibodies compared to the larger AuNP-HRP-E2 conjugates. The cartridge design for E2 detection is divided into three primary zones: detection, bridge, and signal (as detailed in Figure S12). After injection of the incubated sample into the main chamber, the sample flows through the detection zone—functionalized with secondary antibodies—where E2 antibodies bound to either AuNP-HRP-E2 or free E2 are captured. The unbound AuNP-HRP-E2 conjugates, along with CC and FP, then migrate across the bridge to the signal zone, which is functionalized with ChOx. In the signal zone, the enzymatic reactions depicted in Figure 3b take place.

Incubation is a critical step in enhancing signal resolution. Without it, there is insufficient signal differentiation, as AuNP-HRP-E2 is not fully captured in the detection zone (Figure S13). The signal resolution, measured at the difference in REEA signal between plasma samples containing 184 pg/mL and 576 pg/mL of E2, saturated after 15 minutes of incubation, which was optimized as the ideal incubation time for the procedure (Figure 4c).

Another critical parameter in the REEA system is the concentration of AuNP-HRP-E2 conjugates. When the concentration is too high, the binding sites of E2 antibodies during incubation become fully occupied, regardless of the free E2 concentration in the testing solution. To determine the optimal concentration of AuNP-HRP-E2 conjugates, the same paper fluidics setup was used with two testing solutions: one containing 5 ng/mL of free E2 and the other without free E2. The SNR was calculated by dividing the signal from the 5 ng/mL E2 sample by the signal from the sample with no E2, as a function of the AuNP-HRP-E2 conjugate concentration. An SNR of 1 indicates no signal differentiation between the two scenarios. Reducing the AuNP-HRP-E2 concentration corresponding to 0.1 $OD_{525}$ significantly increased the SNR to 5, which was identified as the optimal concentration for the competitive immunoassay (Figure 4d).

Measurement time was optimized to balance signal resolution with detection speed, constrained by the enzymatic reaction duration. The current cartridge design, comprising detection, bridge, and signal generation zones, required 600 seconds to achieve precise signals at concentrations below 5 pg/mL (Figure S14). Notably, the maximum REEA signal was reached within 150–200 seconds (Figure S11), with approximately half of the total measurement time attributed to fluid flow through the detection and signal zones.

Under the optimized testing conditions, the E2-spiked sample was evaluated across a concentration range of 328 fg/mL to 1 ng/mL to determine the LOD of the system (Figure 4e). The LOD was calculated to be approximately 146 fg/mL using the 3-sigma rule, with the CV for each E2 concentration being less than 9.2%. Additionally, clinical plasma samples with E2 concentrations ranging from 19 pg/mL to 4,551 pg/mL were tested (Figure 4f), showing an r² of 0.963 and a standard error of the estimate (Sy/x) of 0.15, as determined by logarithmic regression analysis when compared to results from FDA-cleared clinical immunoassay (Figure 4f).

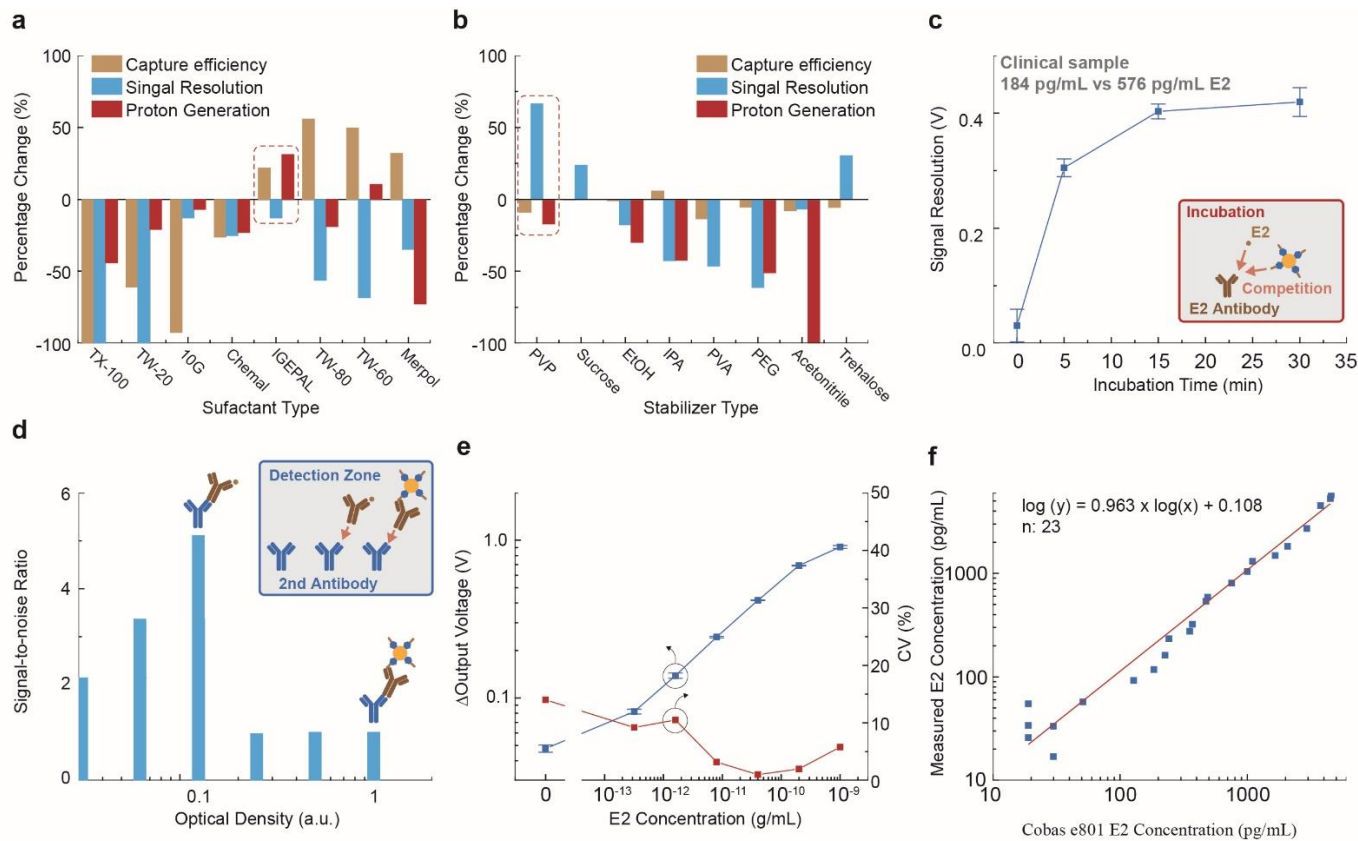

**Figure 4.** Capture efficiency, signal resolution, and proton generation efficiency, with (a) each surfactant and (b) stabilizer mixed into a 2% BSA solution. (c) Signal resolution affected by incubation time. (d) SNR as a function of AuNP-HRP-E2 conjugate concentration in testing samples. (e) $\Delta V_{out}$ and CV values in response to spiked E2 concentrations ranging from 328 fg/mL to 1 ng/mL. (f) Comparison of E2 testing results obtained from the REEA system and an FDA-cleared clinical laboratory immunoassay (Cobas e801, n=23).

## DISCUSSION

This study demonstrates the integration of FET technology with a paper fluidics system via the REEA platform, covering a wide analytical measuring range of E2 concentrations from 19 to 4,551 pg/mL with an r² of 0.963 for clinical plasma samples. By utilizing the FET as a signal transducer, the REEA system achieved a low LOD of 146 fg/mL, effectively addressing the analytical sensitivity limitations typically associated with conventional at-home diagnostic platforms. Conversely, paper fluidics addresses the complexity of FET biosensors, eliminating the need for surface immobilization of biomolecules such as enzymes, DNA, or antibodies. This paper fluidics system also enables low-cost production ($0.55 per test) and offers a long shelf life of up to two years[30].

A key advantage of the REEA platform is its ability to process clinical samples directly, bypassing the need of traditional FET biosensors for capture-release methods that typically require complex washing steps and buffer dilution to address Debye length challenges and matrix effects. This simplification allows for the use of paper-based fluidics, reducing system complexity.

Further development will focus on enhancing usability by enabling the direct application of finger-prick whole blood samples. This will involve integrating a serum separation capability into the paper fluidics to facilitate serum release into the assay cartridge. Pre-drying components such as AuNP-HRP-E2, stabilizers, and surfactants within the paper fluidic system will streamline the workflow, while optimizing paper fluidic structures and antibody screening could eliminate the need for incubation, ensuring signal differentiation with minimal user intervention.

The current detection time of 10 minutes, required for high signal resolution, can be reduced, as the maximum REEA signal is generated in approximately 200 seconds (Figure S11). The lateral paper strip design accounts for approximately 50% of the measurement time, primarily due to fluid passage. A vertical paper fluidics structure could reduce wetting time while maintaining the capture efficiency[31].

Although optimized for E2 detection, the REEA platform has the potential to be expanded for the detection of a broader range of biomarkers, enhancing its versatility in personalized medicine and diagnostics. In line with the global shift towards home-based healthcare, our platform addresses the critical need for accurate, cost-effective, and rapid diagnostics beyond traditional clinical settings. This development represents a significant step towards improving healthcare accessibility, supporting efforts to meet the growing demands on healthcare systems worldwide.

**CONCLUSION**

In conclusion, we have developed a diagnostic platform that integrates REEA with a paper-based fluidic system and a handheld FET reader, enabling proton quantification for biomarker detection. For E2, the platform demonstrated a LOD of 146 fg/mL with a CV below 9.2% in E2-spiked buffer, using a low-cost cartridge ($0.55 per test) within 10 minutes. The REEA platform showed a robust correlation with conventional clinical assays in plasma samples ($r^2$ = 0.963), underscoring its reliability and accuracy. By identifying halogenated phenolic substrates, optimizing buffer components, and incorporating automated $H_2O_2$ generation via an enzymatic ChOx system, the REEA platform shows strong commercialization potential as a cost-effective, accurate, and accessible at-home diagnostic tool. Additionally, it holds significant potential for expansion to a wide array of biomarkers, paving the way for broader clinical and home-based applications.

**AUTHOR CONTRIBUTION**

Electrical measurements and device fabrication were carried out by H.-J. Jang and X. Shi. Interpretation of data and design of platform was carried out by H.-J. Jang, H.-A. Joung, X. Shi, E. Tang, J. Wagner, J. Huang, and J. H. Chen. Simulation was carried out by R. Ding The reader device was designed by B. Ryu and W. Zhuang. The manuscript was prepared by H.-J. Jang, H.-A. Joung, J. Wagner, and J. H. Chen. Plasma samples were retrieved and evaluated by G. Chen and K.-T. J. Yeo. All authors edited the manuscript and commented on it. The project was supervised by J. H. Chen.

**COMPETING INTERESTS**

The authors have a pending patent application related to the technology presented in this study. Dr. Hyun-June Jang and Dr. Hyouarm Joung, co-founders of Kompass Diagnostics, have a commercial interest in the diagnostic technology described in this work.

**ACKNOWLEDGEMENTS**

This work was financially supported by a University of Chicago Faculty Start-up fund and by a Walder Foundation grant

**ADDITIONAL INFORMATION**

Correspondence and requests for materials should be addressed to Junhong Chen.

Supplementary information. The online version contains supplementary material available at

**References**

1  Jones, C. H. & Dolsten, M. Healthcare on the brink: navigating the challenges of an aging society in the United States (vol 10, 22, 2024). *NPJ Aging* **10** (2024).

2  Agyeman-Manu, K. *et al.* Prioritising the health and care workforce shortage: protect, invest, together. *Lancet Glob. Health* **11**, e1162-e1164 (2023).

3  Pandit, J. A., Pawelek, J. B., Leff, B. & Topol, E. J. The hospital at home in the USA: current status and future prospects. *NPJ Digit. Med.* **7** (2024).

4  Liu, W. P. & Lee, L. P. Toward Rapid and Accurate Molecular Diagnostics at Home. *Advanced Materials* **35** (2023).

5  Liu, H., Dao, T. N. T., Koo, B., Jang, Y. O. & Shin, Y. Trends and challenges of nanotechnology in self-test at home. *Trac-Trend Anal. Chem.* **144** (2021).

6  Kersh, E. N., Shukla, M., Raphael, B. H., Habel, M. & Park, I. At-Home Specimen Self-Collection and Self-Testing for Sexually Transmitted Infection Screening Demand Accelerated by the COVID-19 Pandemic: a Review of Laboratory Implementation Issues. *J. Clin. Microbiol.* **59** (2021).

7  Brezina, P. R., Haberl, E. & Wallach, E. At home testing: optimizing management for the infertility physician. *Fertil. Steril.* **95**, 1867-1878 (2011).

8  Wu, A. K., Odisho, A. Y., Washington, S. L., Katz, P. P. & Smith, J. F. Out-of-Pocket Fertility Patient Expense: Data from a Multicenter Prospective Infertility Cohort. *J. Urol.* **191**, 427-432 (2014).

9  Cánovas, R., Daems, E., Langley, A. R. & De Wael, K. Are aptamer-based biosensing approaches a good choice for female fertility monitoring? A comprehensive review. *Biosens. Bioelectron.* **220** (2023).

10  Cauley, J. A. *et al.* Elevated serum estradiol and testosterone concentrations are associated with a high risk for breast cancer. Study of Osteoporotic Fractures Research Group. *Ann. Intern. Med.* **130**, 270-277 (1999).

11  Ankarberg-Lindgren, C. & Norjavaara, E. Estradiol in pediatric endocrinology. *Am. J. Clin. Pathol.* **132**, 978-980 (2009).

12  Li, J. Z. *et al.* Quantitation of estradiol by competitive light-initiated chemiluminescent assay using estriol as competitive antigen. *J. Clin. Lab. Anal.* **34** (2020).

13  Rosner, W., Hankinson, S. E., Sluss, P. M., Vesper, H. W. & Wierman, M. E. Challenges to the Measurement of Estradiol: An Endocrine Society Position Statement. *J. Clin. Endocr. Metab.* **98**, 1376-1387 (2013).

14  Lopreside, A. *et al.* Bioluminescence goes portable: recent advances in whole-cell and cell-free bioluminescence biosensors. *Luminescence* **36**, 278-293 (2021).

15  Na, W., Park, J. W., An, J. H. & Jang, J. Size-controllable ultrathin carboxylated polypyrrole nanotube transducer for extremely sensitive 17beta-estradiol FET-type biosensors. *J. Mater. Chem. B* **4**, 5025-5034 (2016).

16  Jang, H. J. *et al.* Rapid, Sensitive, Label-Free Electrical Detection of SARS-CoV-2 in Nasal Swab Samples. *ACS Appl. Mater. Interfaces* **15**, 15195-15202 (2023).

17  Jang, H. J. *et al.* Remote Floating-Gate Field-Effect Transistor with 2-Dimensional Reduced Graphene Oxide Sensing Layer for Reliable Detection of SARS-CoV-2 Spike Proteins. *ACS Appl. Mater. Interfaces* **14**, 24187-24196 (2022).

18  Jang, H. J. *et al.* Deep Learning-Based Kinetic Analysis in Paper-Based Analytical Cartridges Integrated with Field-Effect Transistors. *ACS Nano* **18**, 24792-24802 (2024).

19  Veitch, N. C. Horseradish peroxidase: a modern view of a classic enzyme. *Phytochemistry* **65**, 249-259 (2004).

20  Sajid, M., Kawde, A. N. & Daud, M. Designs, formats and applications of lateral flow assay: A literature review. *J. Saudi. Chem. Soc.* **19**, 689-705 (2015).

21  Pirzad, R., Newman, J. D., Dowman, A. A. & Cowell, D. C. Horseradish-Peroxidase Assay - Radical Inactivation or Substrate-Inhibition - Revision of the Catalytic Sequence Following Mass-Spectral Evidence. *Analyst* **119**, 213-218 (1994).

22  Sturgeon, B. E., Battenburg, B. J., Lyon, B. J. & Franzen, S. Revisiting the Peroxidase Oxidation of 2,4,6-Trihalophenols: ESR Detection of Radical Intermediates. *Chem. Res. Toxicol.* **24**, 1862-1868 (2011).

23  Gazaryan, I. G. *et al.* Tryptophanless recombinant horseradish peroxidase: Stability and catalytic properties. *Biochem. Bioph. Res. Co.* **262**, 297-301 (1999).

24. Jang, H. J. *et al.* Electronic Cortisol Detection Using an Antibody-Embedded Polymer Coupled to a Field-Effect Transistor. *ACS Appl. Mater. Interfaces* **10**, 16233-16237 (2018).
25. Jang, H. J. *et al.* Analytical Platform To Characterize Dopant Solution Concentrations, Charge Carrier Densities in Films and Interfaces, and Physical Diffusion in Polymers Utilizing Remote Field-Effect Transistors. *J. Am. Chem. Soc.* **141**, 4861-4869 (2019).
26. Pirzad, R., Newman, J. D., Dowman, A. A. & Cowell, D. C. Horseradish-Peroxidase Assay - Optimization of Carbon-Fluoride Bond Breakage and Characterization of Its Kinetics Using the Fluoride Ion-Selective Electrode. *Analyst* **114**, 1583-1586 (1989).
27. Osman, A. M., Boeren, S., Boersma, M. G., Veeger, C. & Rietjens, I. M. Microperoxidase/H2O2-mediated alkoxylating dehalogenation of halophenol derivatives in alcoholic media. *Proc. Natl. Acad. Sci.* **94**, 4295-4299 (1997).
28. Thapa, B. & Schlegel, H. B. Improved pK(a) Prediction of Substituted Alcohols, Phenols, and Hydroperoxides in Aqueous Medium Using Density Functional Theory and a Cluster-Continuum Solvation Model. *J. Phys. Chem. A* **121**, 4698-4706 (2017).
29. Huang, Q. *et al.* Inactivation of horseradish peroxidase by phenoxyl radical attack. *J. Am. Chem. Soc.* **127**, 1431-1437 (2005).
30. O'Farrell, B. Evolution in Lateral Flow-Based Immunoassay Systems. *Lateral Flow Immunoassay*, 1-33 (2009).
31. Ballard, Z. S. *et al.* Deep learning-enabled point-of-care sensing using multiplexed paper-based sensors. *NPJ Digit. Med.* **3** (2020).